# The Design Of Fiber Optic Sensors For Measuring Hydrodynamic Parameters

C.J. Quiett, J.V. Lindesay, D.R. Lyons

Research Center for Optical Physics

Hampton University

## Abstract

We present an approximate analytic model that has been developed for examining hydrodynamic flow near the surface of a fiber optic sensor. An analysis of the conservation of momentum, the continuity equation and the Navier-Stokes equation for compressible flow was used to develop expressions for the velocity $\vec{v}$ and the density $\rho$ as a function of the distance along the flow and above a two dimensional flat surface. When examining the flow near the surface, these expressions can be used to estimate the sensitivity required to perform direct optical measurements and to parameterize the shear force for indirect optical measurements. The use of these results allows for direct modeling of sensors using the optical properties of the flow field. Future work includes analyzing the optical parametric design of fiber optic sensors, modeling sensors to utilize those parameters for optimal measurements involving different mixtures of fluid flow.

## 1    Introduction

We are fabricating fiber optic sensors for hydrodynamic flow applications that could be useful for aerospace and submarine technology. These fiber optic sensors can be used to determine reactant dynamics in harsh environments, where other types of sensors would be unable to make hydrodynamic measurements. Optical fluid flow sensors can be built by embedding fiber optic sensors in durable surfaces, optically measuring the hydrodynamic parameters of compressible fluids flowing over these surfaces. These devices utilize Bragg grating technology to measure the refractive index change of the surrounding medium.

We are not aware of any examination of direct optical measurements of hydrodynamic parameters in the literature. Phillip-Chandy et al[1] designed a fiber optic drag-force flow sensor for measuring the speed and direction of fluid flow, and W. Jin et al[2] has fabricated a fiber optic Bragg grating sensor for measuring flow-induced vibration on a circular cylinder in a cross-flow. In another publication, Denis Roche et al[3] describes a piezoelectric sensor that measures the shear stress in hydrodynamic flow. However, in contrast to electrical sensors, optical sensors are immune to electromagnetic fields and the electrical conduction properties of the environment. Electromagnetic immunity and mechanical and chemical durability are the primary advantages in using fiber optic sensors for measuring hydrodynamic parameters.

## 2     Bragg Sensor History

Hill[4] et al (1978) were the first to actually produce a Bragg reflection grating. The formation of Bragg reflection filters depends upon the photosensitivity of germanium doped fibers in the wavelength region from 170 to 400 nm. This is a particular region of interest since it possesses strong absorption bands for Ge doping [5]. In the original setup, the core of a single mode Ge-doped fiber was exposed to intense laser light at the 488 nm line of argon-ion to establish a standing wave pattern to create an interference filter. Essentially, the formation of the filter is the result of a nonlinear process that produces periodic modulations in the index of refraction within the Ge-doped core region of the fiber. This behavior makes it possible to create very narrow linewidth filters which can then be configured to measure distributed or localized strain as well as do modal mapping along the length of a structure. The ability to produce these grating patterns and the nonlinear mechanisms describing their formation form the basis for ideas involving the use of the length limited Bragg reflection filters, as well as their underlying properties.

Following Hill's initial work, D.R. Lyons repeated their results using the 488 nm line of an argon-ion laser and in an experiment in July 1986 fabricated the first transverse diffraction gratings using a 193 nm KrF excimer laser and subsequently established transverse holographic experimental setups with several UV laser sources using this novel approach[6]. The first experiments to produce the transverse gratings used two interfering UV beams at 193 nm by side illumination of the fiber. This method demonstrated substantial improvement in the fabrication of Bragg gratings and had several advantages. These advantages included lower power requirements to produce interference gratings, the ability to create highly wavelength selective modal discriminators, the capability to write holographic patterns at practically any wavelength above the wavelength of the writing laser, and the inherent facility to write a large number of gratings into a single fiber. Hill's method only permitted a single grating to be written in the fiber at a single wavelength. Later studies have been successful in fabricating Bragg gratings by the technique of side illumination and have been able to extend it to many useful applications[7,8]. An additional application of these gratings is their use in optical electronic multiplexing systems where their wavelength discrimination properties allow large groups of these sensors to be addressed along a single backbone of an optical fiber interconnect[9].

## 3    Modeling of Compressible Subsonic Flow on A Flat Plate

We began modeling the behavior of subsonic compressible flow by examining the conservation of momentum,

$$\int_{A_e} (\rho \vec{v}) \vec{v} \bullet \vec{n} dA = \int_V \rho \vec{g} dV + \int_A \bar{\bar{\tau}}^v \bullet \vec{n} dA \tag{1}$$

the Navier-Stokes equation,

$$\rho \left[ \frac{\partial \vec{v}}{\partial t} + \left( \vec{v} \bullet \vec{\nabla} \right) \vec{v} \right] = \rho \vec{f} - \vec{\nabla} P + \eta \vec{\nabla}^2 \vec{v} + \left( \varsigma + \frac{1}{3} \eta \right) \vec{\nabla} \left( \vec{\nabla} \bullet \vec{v} \right) \tag{2}$$

and the continuity equation,

$$\vec{\nabla} \bullet \rho \vec{v} = 0 \tag{3}$$

We consider a compressible, viscous fluid steadily flowing over a flat plate (see Figure 1).

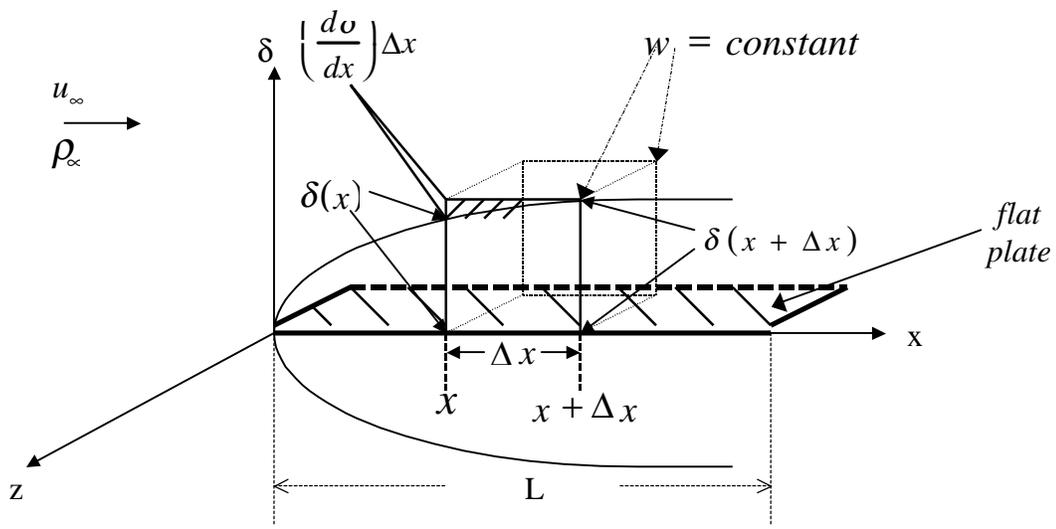

**Figure 1**: Steady, Compressible
Flow Over A Flat Plate

When the fluid encounters the edge of the flat plate, a boundary layer is created near the surface of the plate.  Using the boundary conditions, the momentum conservation equation (Eq. 1) becomes the Boundary Layer Integral-Differential Equation,

$$\frac{d}{dx}\int_0^{\delta(x)}\rho(x,y)v_x^2(x,y)dy - \rho(x,y)u_\infty^2\frac{d\delta}{dx} + \rho(x,\delta(x))u_\infty v_y(x,\delta(x)) = -\eta\left(\frac{\partial v_x(x,y)}{\partial y} + \frac{\partial v_y(x,y)}{\partial x}\right)_{y=0} \quad (4)$$

This equation is used to model compressible subsonic flow on flat plate.  Equations 2 and 3 were used to determine expressions for density and velocity (which are used to calculate the index of refraction change for fluids),

$$n = 1 + \beta\frac{\rho}{\rho_o} \quad (5)$$

and the shear force on the surface,

$$f_x = \int_0^L \tau_{xy}wdx. \quad (6)$$

where $\beta$ is the tabulated refractive density constant, $\rho_0$ is the initial density at normal conditions, $\rho$ is the density at the measurement position, $w$ is the width of the flat plate, and $L$ is the length of the sensor.

# 4 Expansion of Hydrodynamic Equations Near Flat Surfaces

The expressions for $v_x$, $v_y$, and $\rho$ below are represented as polynomials of $\left(\frac{y}{\delta}\right)$, where the boundary layer thickness $\delta$ is a function of the distance along the flow direction $x$:

$$v_x(x,y) = u_\infty \left(\frac{y}{\delta}\right) \left[ (1 - m - q) + m\left(\frac{y}{\delta}\right) + q\left(\frac{y}{\delta}\right)^2 \right]$$

$$v_y(x,y) = \left[ f + g\left(\frac{y}{\delta}\right) + h\left(\frac{y}{\delta}\right)^2 \right] \frac{d\delta}{dx} \qquad (7)$$

$$\rho(x,y) = \rho_\infty \left[ (1 - a - b) + a\left(\frac{y}{\delta}\right) + b\left(\frac{y}{\delta}\right)^2 \right]$$

These expressions are formulated to satisfy the following boundary conditions:

1. velocity vanishes at the surface due to viscous sticking ($v_x = 0$ and $v_y = 0$ for y=0),

2. flow becomes laminar outside of the boundary layer ($v_x = u_\infty$ and $v_y \approx 0$ when $y = \delta$), and

3. stress should be negligibly small at the boundary layer interface ($\tau_{yx}^v \approx 0$).

The flow parameters [Eqs. (7)] can be substituted into the Navier-Stokes, and powers of $\left(\frac{y}{\delta}\right)$ examined to solve for coefficients within each expression. This gives a form for which will be discussed in the next section. The scales of the physical parameters are related by taking note of the following:

1) $v_s^2 = \frac{\partial P}{\partial \rho}$, speed of sound squared (related to the compressibility), $\therefore$,

$$\frac{\partial P}{\partial x} \cong -a v_s^2 \rho_\infty \left(\frac{y}{\delta^2}\right)\frac{\partial \delta}{\partial x} \qquad \text{and} \qquad \frac{\partial P}{\partial y} \cong v_s^2 \rho_\infty \left[ a\left(\frac{1}{\delta}\right) + 2b\left(\frac{y}{\delta^2}\right) \right],$$

2) $\frac{\partial v_x}{\partial y} \ll \frac{\partial v_y}{\partial x}$, this implies that $\left(\frac{y}{\delta}\right)\frac{\partial \delta}{\partial x} \ll 1$.

The zeroth and first orders of the x and y components were examined to solve for the variables

$\Rightarrow$ x-component

- zeroth order

$$\frac{\rho_\infty u_\infty}{\delta} f(1-a-b)(1-m-q)\frac{d\delta}{dx} = \frac{2m\eta u_\infty}{\delta^2} \tag{8a}$$

- first order

$$\rho_\infty u_\infty \left\{ (1-a-b)\left[ g(1-m-q) + 2fm \right] \right\} \left( \frac{y}{\delta^2} \right)\frac{d\delta}{dx} = av_s^{\ 2}\rho_\infty \left( \frac{y}{\delta^2} \right)\frac{d\delta}{dx} + 6q\eta u_\infty \left( \frac{y}{\delta^3} \right) \tag{8b}$$

$\Rightarrow$ y-component

- zeroth order

$$0 = -av_s^{\ 2}\rho_\infty \left( \frac{1}{\delta} \right) + \left\{ 2h\eta + \left( \zeta + \frac{1}{3}\eta \right)\left[ 2h - u_\infty(1-m-q) \right] \right\} \left( \frac{y}{\delta^2} \right)\frac{d\delta}{dx} \tag{9a}$$

- first order

$$0 = -2bv_s^{\ 2}\rho_\infty \left( \frac{y}{\delta^2} \right) + \left( \zeta + \frac{1}{3}\eta \right)\left[ -4u_\infty m \left( \frac{y}{\delta^3} \right)\frac{d\delta}{dx} \right]. \tag{9b}$$

The coefficients can then be substituted into the expressions for $v_x$, $v_y$, and $\rho$ [Eqs. (7)], which become:

$$\rho \approx \rho_\infty \left[ 1 + \frac{\frac{3}{2}u_\infty\eta}{\rho_\infty v_s^{\ 2}} \left( \frac{y}{\delta^2} - \frac{1}{\delta} \right)\frac{d\delta}{dx} \right]$$

$$v_x \approx u_\infty \left[ \frac{3}{2}\left( \frac{y}{\delta} \right) - \frac{1}{2}\left( \frac{y}{\delta} \right)^3 \right] \tag{10}$$

$$v_y \approx u_\infty \left[ \frac{3}{4}\left( \frac{y}{\delta} \right)^2 - \frac{3}{8}\left( \frac{y}{\delta} \right)^4 \right]\frac{d\delta}{dx}$$

For continuity, the higher order terms in Eq. (10) have been included.

# 5 Developing the Compressible Boundary Layer Thickness Function

Equations (10) can be substituted into the boundary layer integral-differential equation to determine the form of the compressible boundary layer thickness function $\delta(x)$ on a flat plate. Equation 11 gives the boundary layer integral-differential equation [Eq. (6)] that includes the expressions for $v_x$, $v_y$, and $\rho$.

$$\frac{477}{2240}\frac{u_\infty^3 \eta}{v_s^2}\frac{d^2\delta}{dx^2} + \frac{39}{280}\rho_\infty u_\infty^2 \frac{d\delta}{dx} = \frac{3/2\,\eta u_\infty}{\delta} \qquad (11)$$

From this non-linear, second order differential equation, the form of the compressible boundary layer thickness function can be derived

$$\delta(x) \cong \sqrt{\frac{280}{13}\Lambda x}\left[1 - \frac{1043}{1456}\frac{\Lambda}{x}\left(\frac{u_\infty}{v_s}\right)^2\right], \qquad (12)$$

valid when $\frac{\Lambda}{x} << 1$, where $\Lambda \equiv \frac{\eta}{\rho_\infty u_\infty}$. Equations (10) were further reduced by substituting the form of $\delta(x)$ in each expression. The final expressions for $v_x$, $v_y$, and $\rho$ are

$$\rho \approx \rho_\infty\left[1 - \frac{3}{4}\left(\frac{u_\infty}{v_s}\right)^2\left(1 - \frac{y}{\delta}\right)\frac{\Lambda}{x}\right]$$

$$v_x \approx u_\infty\left[\frac{3}{2}\left(\frac{y}{\delta}\right) - \frac{1}{2}\left(\frac{y}{\delta}\right)^3\right] \qquad (13)$$

$$v_y \approx u_\infty\sqrt{\frac{70}{13}\frac{\Lambda}{x}}\left[\frac{3}{4}\left(\frac{y}{\delta}\right)^2 - \frac{3}{8}\left(\frac{y}{\delta}\right)^4\right]\left[1 + \frac{1043}{1456}\frac{\Lambda}{x}\left(\frac{u_\infty}{v_s}\right)^2\right]$$

# 6 Geometrical Modeling and Performance Predictions

In a general sensor design, a gas such as nitrogen will be flowed over an embedded Bragg sensor (see Figure 2).

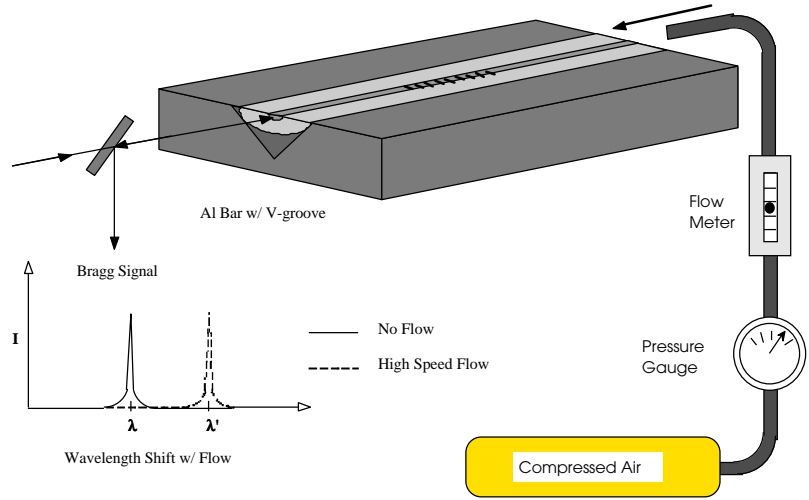

**Figure 2** Optical flow sensor setup

Since the fiber optic sensor is located near the surface (recall Figure 1), the expression for density when $y = 0$ is

$$\rho(x,0) \approx \rho_\infty \left[ 1 - \tfrac{3}{4} \left( \frac{u_\infty}{v_s} \right) \frac{\Lambda_o}{x} \right], \tag{14}$$

where $\Lambda_o \equiv \dfrac{\eta}{\rho_\infty v_s}$ (recall, $\Lambda \equiv \dfrac{\eta}{\rho_\infty u_\infty}$). Equation (14) was used to calculate the refractive index change for nitrogen. From Eq. (4), the equation for the index of refraction becomes

$$\frac{\delta n}{n} \cong \beta \frac{\delta \rho}{\rho} \approx -\tfrac{3}{4} \beta \frac{\delta u_\infty}{v_s} \frac{\Lambda_o}{x}. \tag{15}$$

Therefore the refractive index change for nitrogen (for example) is given by

$$\frac{\delta n}{n} \approx -9.70 \times 10^{-9} \left( \frac{\delta u_\infty}{v_s} \right) \tag{16}$$

$$\eta = 17.9 \times 10^{-6} \, N\!/\!m \text{ @ 26°C for N}_{2\text{(g)}}$$

$$\beta = 3 \times 10^{-4} \text{ for N}_{2\text{(g)}}$$

where $\rho_\infty = 1.21 \, kg\!/\!m^3$ @ 20°C for air and $\Lambda_o = 4.31 \times 10^{-8} \, m$.

$$\upsilon_s = 343 \, m\!/\!s \text{ @ 20°C for air}$$

$$x = 1 \times 10^{-3} \, m$$

These quantities are values chosen for a typical gas sensor design. For a direct optical measurement, the refractive index change for the compressible flow can be calculated to be of the order of $10^{-9}$ for air. For liquids, the index change will be considerably larger. The constant $\Lambda_o$ is seen to provide a natural length scale for fluid flow properties.

In a similar setup (see Figure 3), a shear stress fiber optic sensor can be developed by attaching a fiber to two flat plates.

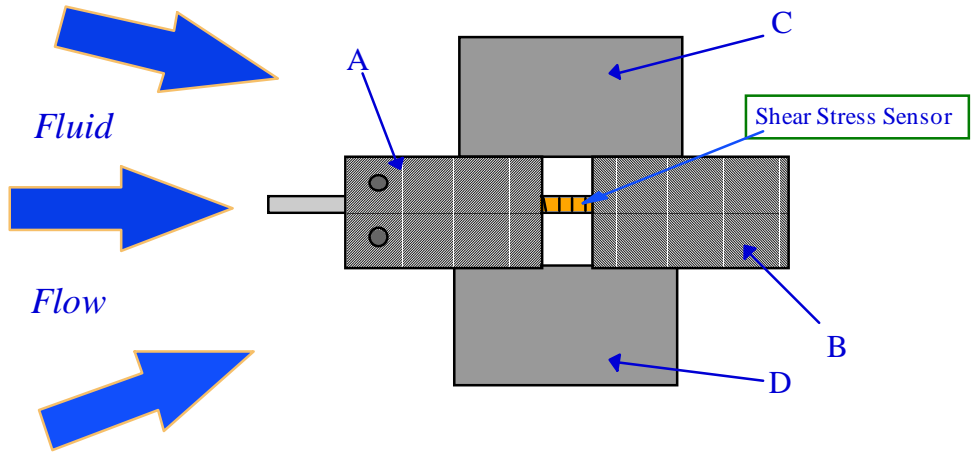

**Figure 3** Bragg shear stress sensor setup (gap exaggerated)

For this shear stress sensor, the total stress is

$$\tau_{xy} \approx \frac{-\tfrac{3}{2} u_\infty \eta}{\left\{ \sqrt{\frac{280}{13} \Lambda x} \left[ 1 - \frac{1043}{1456} \frac{\Lambda}{x} \left( \frac{u_\infty}{c} \right)^2 \right] \right\}}. \qquad (17)$$

$$\therefore, \ f_x \approx -\frac{3}{2}\rho_\infty u_\infty^2 w \sqrt{\frac{13}{70}\Lambda L}, \tag{18}$$

where $\Lambda \equiv \Lambda_o\left(\dfrac{v_s}{u_\infty}\right) \ll L$. Using the dimensions of a prototype device,

$$w = 3\text{cm}$$
$$L = 3\text{cm}$$

the flux of the stress tensor for air flow is be given by

$$f_x \approx -9.93 \times 10^{-2}\left(\frac{u_\infty}{v_s}\right)^{\frac{1}{2}} \text{Newtons}$$

relative to the speed of sound, $v_s$. This can be related with the tensile strength of the fiber to determine wavelength shifts in a Bragg filter corresponding to particular fluid flows.

## 7     Conclusion

An approximate analytic model for examining the hydrodynamic parameters near the surface of a fiber optic sensor has been developed. In terms of the parameters, the sensitivity required to perform direct optical measurements can be estimated for the refractive index change.

The expressions for the density and the velocity (in two dimensions) were represented as polynomials of $\left(\dfrac{y}{\delta}\right)$, where $\delta$ is a function of $x$. The modeling of these parameters can be used to design surfaces which are optimal for measuring the flow properties within the boundary layer such as viscosity, compressibility, and pressure. Fiber optic sensors are being designed to optically measure the change in density and velocity changes using these results. Although results for liquid flow are seen to be significant, the flow dependence of optical parameters in gases is seen to be small. It is hopeful that clever experimental design utilizing, for instance, mode mixing or interferometry will result in sensors which can detect minute flows in the optical properties of hydrodynamic flows.

### Acknowledgements

The authors would like to acknowledge the support of the Fiber Optic Sensors and Smart Structures group in the Research Center for Optical Physics at Hampton University. Related aspects of this work are described in Masters of Science thesis (2000) for CQ.